\documentclass[lettersize,journal]{IEEEtran}

\usepackage{amsmath,amsfonts}
\usepackage{array}
\usepackage[caption=false,font=normalsize,labelfont=sf,textfont=sf]{subfig}
\usepackage{textcomp}
\usepackage{stfloats}
\usepackage{url}
\usepackage{verbatim}
\usepackage{graphicx}
\usepackage{cite}
\usepackage[linesnumbered,ruled,vlined, noend]{algorithm2e}
\usepackage{algorithmic}
\usepackage[T1]{fontenc}

\usepackage{pifont}
\usepackage{makecell}
\usepackage{tablefootnote}

\usepackage{bm}
\usepackage{xfrac}
\usepackage{tabularx}
\usepackage{multirow}
\usepackage[multiple]{footmisc}
\usepackage{enumitem}
\usepackage{xcolor, colortbl}

\usepackage[hidelinks]{hyperref}
\usepackage{cleveref}
\crefformat{section}{\S#2#1#3}
\crefformat{subsection}{\S#2#1#3}
\crefformat{subsubsection}{\S#2#1#3}
\crefname{equation}{Equation}{Equations}

\usepackage{booktabs}
\usepackage{footnote}
\usepackage{tikz}
\usepackage{adjustbox}

\begin{document}
\bstctlcite{IEEEexample:BSTcontrol}

\title{ViTrail: Transmission--Inference Co-Design for Vision Transformer--Based Mobile Video Analytics}

\author{Miao~Zhang,~\IEEEmembership{Member,~IEEE,} Guanzhen~Wu,~\IEEEmembership{Graduate Student Member,~IEEE,} Hao~Fang, Yifei~Zhu,~\IEEEmembership{Member,~IEEE,} Fangxin~Wang,~\IEEEmembership{Member,~IEEE,} Ruixiao~Zhang, Linfeng~Shen, and Jiangchuan~Liu,~\IEEEmembership{Fellow,~IEEE}
\thanks{Miao Zhang, Guanzhen Wu, Hao Fang, Linfeng Shen, and Jiangchuan Liu are with the School of Computing Science, Simon Fraser University, Canada. Yifei Zhu is with Global College, Shanghai Jiao Tong University, China. Fangxin Wang is with School of Science and Engineering (SSE) and The Future Network of Intelligence Institute (FNii), The Chinese University of Hong Kong, Shenzhen, China. Ruixiao Zhang is with Computer Science Department, University of Illinois Urbana-Champaign, United States.}}

\maketitle

\begin{abstract}
Edge-assisted mobile video analytics (MVA) relies on a tight device--edge feedback loop in which captured frames are transmitted for remote inference and the returned results are rendered on the mobile device. Many existing systems optimize frame transmission while treating server-side inference as a fixed cost. This separation becomes inefficient for vision transformer (ViT)-based dense prediction, where high-resolution inputs simultaneously increase network traffic and server computation. We present \textbf{ViTrail}, which co-designs transmission and ViT inference through a shared mixed-resolution representation. The device preserves task-relevant regions at full resolution and downsamples less informative regions before encoding. The server tokenizes each region at its transmitted resolution, so the same regional resolution assignment reduces both the upload payload and the number of tokens processed by the ViT. For dense prediction, ViTrail aligns regions with ViT patches and attention windows and restores a regular feature grid at a configurable stage. At runtime, it selects regional downsampling, compression quality, and the restoration point to maximize predicted installation accuracy, the accuracy of the rendered result when a new remote result is first used for rendering. We build a fully functional prototype on commodity mobile and edge devices and evaluate it in real time across 300 video--trace pairs with real-world 4G/5G traces. ViTrail achieves the highest mean rendering accuracy among the evaluated systems and leads each baseline on 81\% to 95\% of pairs, while reducing median end-to-end offloading latency by 38\% against the most accurate baseline.
\end{abstract}

\begin{IEEEkeywords}
Mobile video analytics, vision transformers, mixed-resolution offloading
\end{IEEEkeywords}

\section{Introduction}

\begin{figure}[!t]
    \centering
    \includegraphics[width=0.95\linewidth]{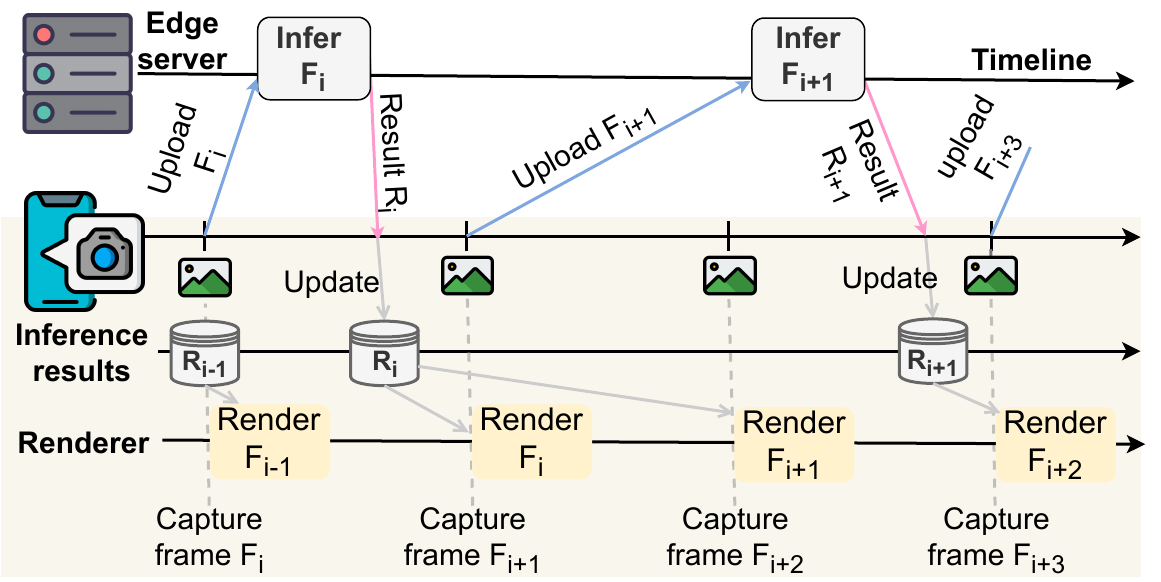}
    \caption{Edge-assisted mobile video analytics.}
    \label{fig:mva}
    \vspace{-0.5cm}
\end{figure}

\IEEEPARstart{M}{obile} video analytics (MVA) supports applications ranging from augmented reality (AR)~\cite{meng2022we, kong2023accumo, kong2024arise} to autonomous mobile sensing. These applications rely on accurate vision models, whose computational demands often exceed the resource budgets of mobile devices. The \emph{edge-assisted paradigm} therefore offloads captured frames to an edge server for inference and returns the results to the device for rendering, such as overlaying bounding boxes or labels on the live camera view~\cite{chen2015glimpse, liu2019edge, zhang2021elf}. As Fig.~\ref{fig:mva} illustrates, this process forms a latency-sensitive feedback loop comprising device-side encoding, uplink transmission, server-side decoding and inference, and result delivery. Consequently, rendering quality depends on both remote inference accuracy and result freshness. When end-to-end (E2E) offloading latency is high, newly captured frames must be rendered using older remote results. A lightweight local tracker can propagate these results to newer frames, but tracking drift and scene changes can degrade rendering accuracy as the results age~\cite{liu2019edge, meng2022we, kong2023accumo}.

Within this feedback loop, the choice of vision backbone directly affects server-side inference delay. While prior edge-assisted MVA systems have predominantly relied on convolutional neural network (CNN) backbones, vision transformers (ViTs)~\cite{dosovitskiy2020image} can model global context and have demonstrated strong performance in dense prediction tasks~\cite{strudel2021segmenter, li2022exploring, xu2022vitpose}. These tasks require high-resolution inputs to preserve fine-grained spatial information~\cite{zhang2021elf}, and the token count and inference cost of a ViT grow with the input resolution even under window attention. In our measurements, ViTDet-L~\cite{li2022exploring}, a representative ViT-based dense prediction model employing window attention, requires about 265\,ms to process a 1080p frame on an edge server equipped with an NVIDIA RTX 5090 GPU, equivalent to approximately 3.8 frames per second (FPS) for sequential inference. Server-side inference can therefore become a major component of E2E offloading latency.

This changed latency profile calls for joint optimization of transmission and inference, which existing techniques address only in part. Transmission-oriented MVA techniques reduce upload traffic through mixed-quality compression and region masking~\cite{liu2019edge, liu2022adamask, chen2022context}. Both operations preserve the server-side input dimensions, leaving the ViT token count and inference cost unchanged. Model-centric ViT acceleration methods prune, merge, or reorganize tokens~\cite{kong2022spvit, fayyaz2022adaptive, haurum2023tokens, ronen2023vision, havtorn2023msvit}. Applied on the edge server, they reduce model computation only after the full frame has been transmitted. Uniformly downsampling the entire frame reduces both costs~\cite{zhang2017live, zhang2018awstream, zhang2022casva}, but it treats task-relevant and less informative regions alike and can remove the fine-grained details required for accurate dense prediction.

Prior work has explored mixed-resolution tokenization to reduce ViT computation in image classification and segmentation~\cite{ronen2023vision, havtorn2023msvit}. Our key insight is that choosing regional resolutions on the mobile device before encoding lets the same choice govern both the upload size and the number of input tokens processed at the edge. The device preserves task-relevant regions at full resolution and downsamples less informative regions, and this mixed-resolution representation leaves fewer pixels to encode. The edge server then tokenizes each region at its transmitted resolution, so downsampled regions contribute fewer tokens to the ViT.

Turning this insight into a practical MVA system raises two challenges. First, regional downsampling must preserve alignment with the backbone's attention windows, and the server must recover the regular feature grid required by the dense prediction head from tokens at different resolutions. Second, the system must jointly configure transmission and inference at runtime for each offloaded frame, because task-relevant regions and uplink throughput change over time. Regional resolution, compression quality, and the stage at which the server restores the regular feature grid together affect the fidelity of a result and the delay before it can be used for rendering. A configuration with higher inference accuracy therefore does not always yield higher rendering accuracy, because its result may return to the mobile device later and be more stale when it is rendered.

To address these challenges, we present \textbf{ViTrail}, a ViT-native system for edge-assisted MVA. Our contributions are as follows:

\begin{itemize}[leftmargin=*]
    \item[$\circ$] \textbf{Transmission--inference co-design.}
    We design a shared mixed-resolution representation that reduces both the upload size and the initial ViT token count. The mobile device downsamples less informative regions before encoding, and the edge server tokenizes each region directly at its \mbox{transmitted} resolution.

    \item[$\circ$] \textbf{Mixed-resolution processing for dense prediction.}
    We make mixed-resolution inputs compatible with both the ViT backbone and the dense prediction head. Each region aligns with ViT patches and attention windows whether it is kept at full resolution or downsampled. We restore the regular feature grid required by the head at a configurable backbone stage. Adjusting this stage controls the trade-off between inference cost and accuracy.

    \item[$\circ$] \textbf{Installation-accuracy-driven adaptation.}
    We introduce installation accuracy, the accuracy of the rendered result when a new remote result is first used for rendering, as the configuration-selection objective rather than inference accuracy alone. ViTrail combines content- and network-aware latency estimates with age-dependent accuracy profiles to select the configuration that maximizes predicted installation accuracy under a freshness target.

    \item[$\circ$] \textbf{Prototype implementation and evaluation.}
    We build a fully functional prototype on commodity mobile and edge devices and evaluate it in real time across 300 video--trace pairs with real-world 4G/5G traces. In head-to-head comparisons with the four baselines, ViTrail achieves higher rendering accuracy on 81\% to 95\% of the pairs. It also reduces median E2E offloading latency by 38\% relative to AdaptRQ, the baseline with the highest mean rendering accuracy. Ablation and sensitivity analyses further confirm the contribution of the key design choices and the robustness of ViTrail.
\end{itemize}

\section{Background and Related Work}
\label{sec:background}

\subsection{Edge-Assisted Mobile Video Analytics}
Edge-assisted MVA offloads compute-intensive vision inference to edge servers while using local processing to maintain responsive rendering~\cite{chen2015glimpse}. Liu et al.~\cite{liu2019edge} decouple rendering from offloading and combine dynamic region-of-interest (RoI) encoding with motion-vector-based tracking. Other systems reduce upload traffic through server-driven or content-aware compression~\cite{du2020server,liu2022adamask,chen2022context}. Elf~\cite{zhang2021elf} partitions frames around region proposals and distributes inference across multiple edge servers. AccuMO~\cite{kong2023accumo} and ARISE~\cite{kong2024arise} coordinate offloading across tasks or clients, while FPSelector~\cite{zhu2025fpselector} selects network paths for mobile AR traffic. In the broader setting of edge--cloud video analytics, ModelIO~\cite{wang2022modelio} selects the inference location, detection model, and input resolution for each frame. ViTrail instead assigns resolutions per region before encoding.

\subsection{ViT Backbones for Dense Prediction}
A vanilla ViT divides an input image into fixed-size, non-overlapping patches, projects them into token embeddings, adds positional embeddings, and processes the resulting token sequence using a Transformer encoder~\cite{dosovitskiy2020image}. The feature map remains \emph{single-scale} across encoder blocks, while global self-attention captures long-range dependencies across the image. Dense prediction tasks in MVA, such as object detection and semantic segmentation, require fine-grained spatial information and therefore typically operate on high-resolution inputs. Image classification, by comparison, commonly uses inputs of 224$\times$224 pixels~\cite{caron2021emerging, he2022masked}. For global self-attention, computation scales quadratically with the input sequence length, making high-resolution inputs expensive~\cite{havtorn2023msvit}. Even with window attention, our profiling shows that increasing the input resolution from 360p to 1080p increases ViTDet-L inference delay by 7.2$\times$, motivating reductions in the number of tokens processed during inference.

\subsection{Inference-Time Acceleration for ViTs}
Since the token sequence need not have a fixed length, inference can trade accuracy for computation by dynamically varying the number and spatial arrangement of tokens, with appropriate positional encoding~\cite{dosovitskiy2020image}. Existing methods reduce visual tokens through two broad strategies. One starts from a uniform patch grid and shortens the sequence through early patch selection~\cite{hu2025sampl}, intermediate-layer token pruning~\cite{rao2021dynamicvit,fayyaz2022adaptive,long2023beyond}, or token merging~\cite{bolya2023token}. Another changes the tokenization itself using adaptive or mixed-resolution tokenization~\cite{ronen2023vision,havtorn2023msvit}, assigning more tokens to task-relevant regions and fewer to less informative regions.

\begin{figure*}
  \centering
  \includegraphics[width=0.95\linewidth]{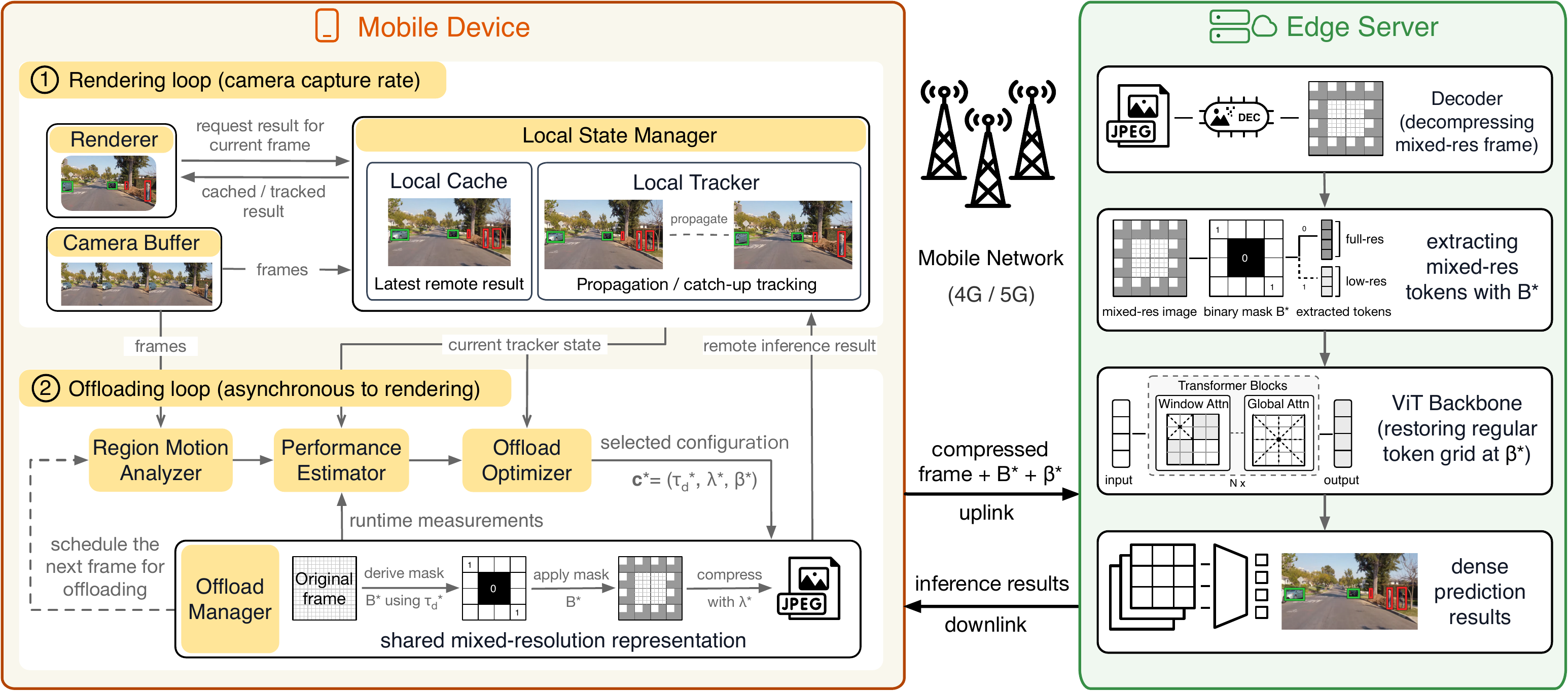}
  \caption{System overview of ViTrail.}
  \label{fig:system-overview}
\end{figure*}

Many token-reduction methods were designed for classification, where the prediction head can operate on a global representation. Token pruning leaves gaps in the spatial grid, while token merging and mixed-resolution tokenization produce tokens that cover regions of different sizes. Applying a conventional dense prediction head to these tokens therefore requires reconstructing a spatially aligned feature grid.

Model splitting, a separate line of work, distributes ViT computation across machines, as in SPViT~\cite{zhao2025spvit} and LVMScissor~\cite{liu2025lvmscissor}. Janus~\cite{jiang2025janus} adapts both model splitting and token pruning to network conditions. ViTrail runs the entire ViT on the edge server. With mixed-resolution tokenization, downsampling selected regions before encoding reduces both the uplink payload and the server's initial token count.

\section{ViTrail System Overview}

We first present an overview of ViTrail, a ViT-native MVA system built around a regional resolution assignment shared by mobile transmission and edge inference. As shown in Fig.~\ref{fig:system-overview}, ViTrail follows the canonical edge-assisted MVA paradigm and organizes its device-side operation into two loops that run at different rates. A rendering loop produces an analytics result for every captured frame, while an offloading loop initiates a new remote update whenever the preceding offload completes. The two loops are coupled through the remote results that the offloading loop obtains and the rendering loop consumes.

The rendering loop runs at the camera capture rate. The device camera continuously places captured frames in a Camera Buffer, and the Renderer queries the Local State Manager at the same rate for the analytics result associated with the frame to be displayed. Because the E2E offloading latency often exceeds the frame capture interval, such as 33\,ms at 30\,FPS, the remote result for the current frame is frequently unavailable by its rendering deadline. The Local State Manager therefore maintains a Local Cache containing the latest available remote inference result and a Local Tracker that propagates this result to subsequent frames, helping maintain rendering accuracy when fresh remote results are unavailable~\cite{chen2015glimpse,kong2023accumo,liu2019edge}. When queried, it returns the cached remote result if it corresponds to the requested frame and otherwise returns the tracker-propagated result.

The offloading loop selects and applies a configuration for each frame scheduled for remote inference. We define a configuration $\bm{c}\in\mathcal{C}$ as $\bm{c}=(\tau_d,\lambda,\beta)$, where $\mathcal{C}$ is the set of candidate configurations, $\tau_d$ specifies a regional downsampling policy, $\lambda$ controls the compression quality of the resulting mixed-resolution frame, and $\beta$ specifies the restoration point, the backbone stage at which low-resolution tokens are restored to a regular spatial grid, as detailed in \cref{sec:restore-pos}. The Region Motion Analyzer and Local Tracker supply the motion and task relevance of each decision region in the current frame, which determine the downsampling mask of each candidate policy. The Performance Estimator combines a compressed-size prediction, profiled processing delays, and recent network observations to estimate each candidate's E2E offloading latency. It adds an online estimate of the catch-up lag to predict the installation age, the age of the result when it is first used for rendering, and obtains the installation accuracy at that age from offline accuracy profiles. The Offload Optimizer then selects $\bm{c}^\star$ by maximizing the predicted installation accuracy subject to the freshness constraint described in \cref{sec:config-select}.

Given $\bm{c}^\star$, the Offload Manager applies the selected regional downsampling policy $\tau_d^\star$ to derive a frame-specific binary downsampling mask $\mathbf{B}^\star$. Each entry indicates whether the corresponding decision region is downsampled. The Offload Manager constructs a mixed-resolution frame using $\mathbf{B}^\star$ and compresses it with $\lambda^\star$. The compressed frame, $\mathbf{B}^\star$, and $\beta^\star$ are then transmitted to the server. After decoding, the server uses $\mathbf{B}^\star$ to construct the mixed-resolution token sequence and restores the regular token grid at the backbone stage specified by $\beta^\star$. The returned result completes the current offloading cycle. The Offload Manager forwards it to the Local State Manager, which performs catch-up tracking asynchronously to refresh the cached and tracking states. Until catch-up tracking completes, rendering continues from the previous remote result. The remainder of the paper uses object detection to illustrate ViTrail's key system designs.

\section{Shared Mixed-Resolution Representation}

\begin{figure}[!t]
    \centering
    \includegraphics[width=\linewidth]{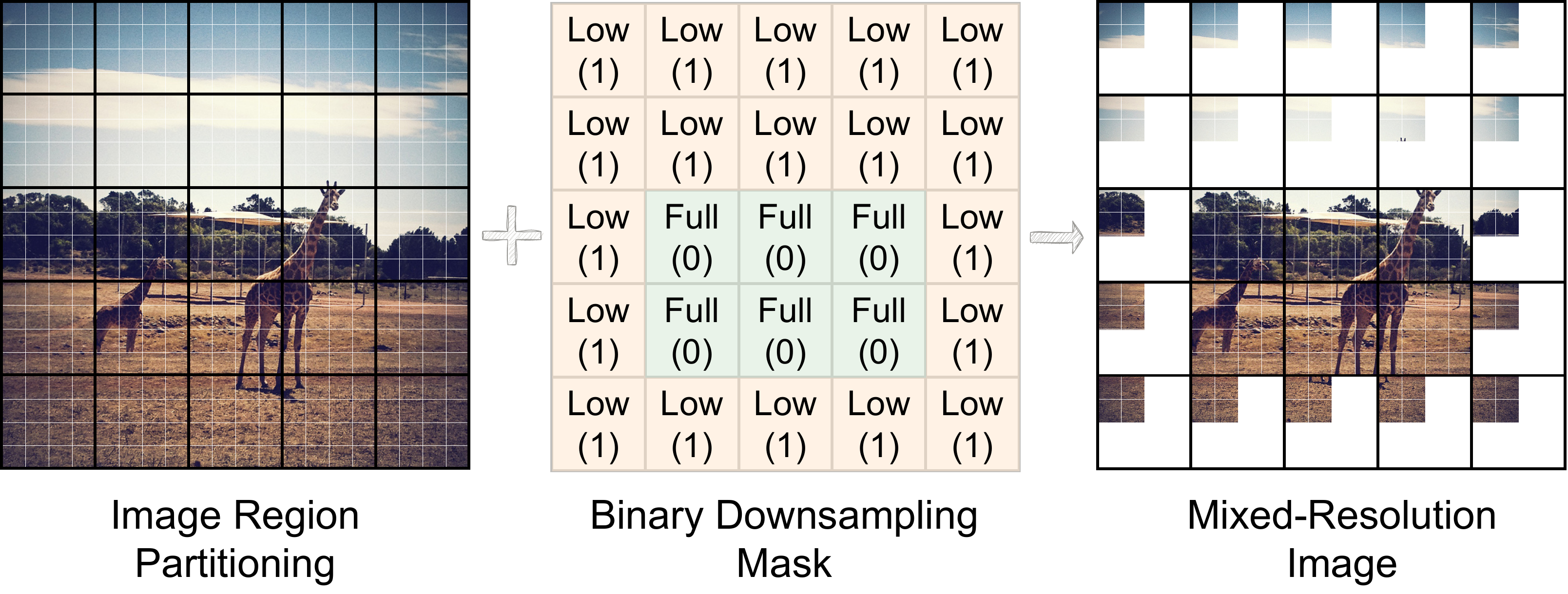}
    \caption{Inference-compatible region partitioning.}
    \label{fig:region-partitioning}
    \vspace{-0.5cm}
\end{figure}

The mixed-resolution frame serves as the common representation for transmission and inference. The device uploads it, and the server directly tokenizes its full- and low-resolution regions, so the same regional resolution assignment governs both the encoded payload and the initial token sequence length. This coupling introduces two compatibility requirements. The input representation must align with the ViT backbone's patchification and window-attention partitions (\cref{sec:region-partition}), while a regular, spatially aligned token grid must be recovered before dense prediction (\cref{sec:restore-pos}).

\subsection{Inference-Compatible Region Partitioning}
\label{sec:region-partition}

Meeting the input-side constraint requires choosing the spatial granularity at which downsampling decisions are made. ViT backbones tokenize images into a regular grid of fixed-size patches, typically 16$\times$16 pixels each~\cite{dosovitskiy2020image}, while backbones designed to handle high-resolution inputs often restrict self-attention in intermediate Transformer blocks to non-overlapping $w \times w$ token windows~\cite{li2022exploring,ryali2023hiera}. Decision regions must therefore align with the native patch grid and be sized so that their full- and low-resolution token grids each decompose into complete attention windows. Otherwise, region boundaries cut across the fixed window partitioning assumed by the backbone.

To address this issue, we propose an \emph{inference-compatible region partitioning} method that aligns with the patchification and window-attention mechanisms of pre-trained ViT backbones. An illustrative example is shown in Fig.~\ref{fig:region-partitioning}, where a coarse-grained grid (shown in black) is overlaid on the image. Each black grid cell, termed a \emph{decision region}, spans $r \times r$ image patches, with each patch (shown in white) being 16$\times$16 pixels. Downsampling decisions are made at the decision region level, producing a binary region-level downsampling mask. The regions marked as low resolution are then downsampled by a factor of $d$, reducing the number of image patches (and thus tokens) required to represent these regions by a factor of $d^2$. To ensure compatibility with the ViT backbone's window-attention computation, we set $r = w d$, where $w$ is the window size used by the ViT backbone to calculate window attention. In the example shown, both $w$ and $d$ are set to 2, yielding $r=4$. This choice ensures that, within each decision region, the full-resolution tokens can be partitioned into complete $w\times w$ attention windows, while the downsampled tokens form exactly one $w\times w$ window.

Because $r = wd$, the factor $d$ also sets how finely regions can be selected. A larger $d$ removes more tokens from each downsampled region but requires larger decision regions, so keeping a small object at full resolution also keeps more of the background around it. We therefore use $d=2$, the smallest integer factor above one, which yields the finest regions and still reduces each downsampled region's tokens to one quarter.

\subsection{Flexible Resolution Restoration}
\label{sec:restore-pos}

Dense prediction heads typically consume multi-scale feature maps, whereas vanilla ViT backbones produce a single-scale feature map. Prior work~\cite{li2022exploring} bridges this gap by constructing a feature pyramid from the single-scale output, but this construction assumes that the tokens uniformly tile the image. Mixed-resolution tokenization violates this assumption. As Fig.~\ref{fig:region-partitioning} shows, a token from a region downsampled by a factor of $d$ represents $d^2$ times the image area of a full-resolution token. Consequently, the mixed-resolution token sequence cannot be directly reshaped into a regular 2D feature grid. A restoration mechanism is therefore required before a spatially aligned feature pyramid can be constructed.

\begin{figure*}
    \centering
    \includegraphics[width=0.95\linewidth]{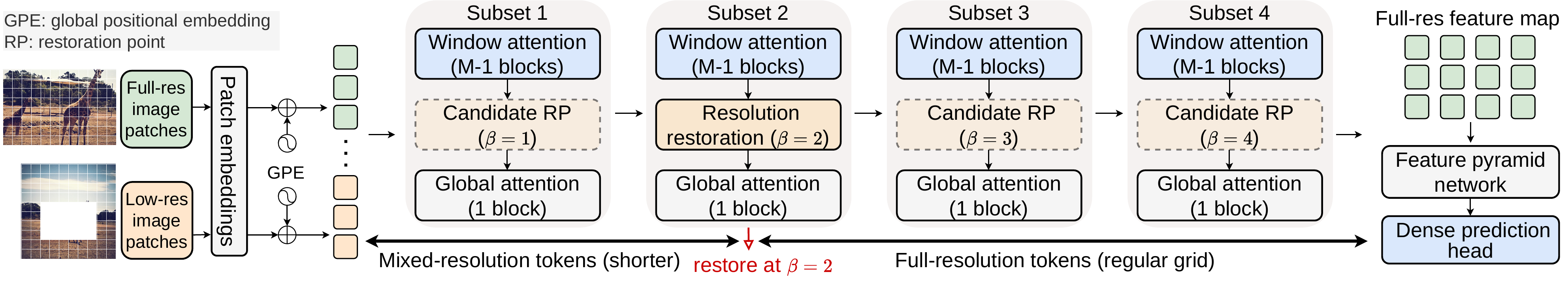}
    \caption{Flexible resolution restoration in ViTrail, illustrated using ViTDet-L with four Transformer subsets and \(\beta=2\).}
    \label{fig:mix-res-vitdet}
    \vspace{-0.3cm}
\end{figure*}

To address this challenge, we propose a configurable resolution-restoration mechanism that restores the resolution of token features associated with downsampled regions at a selected stage within the ViT backbone. A common organization for ViT backbones designed to handle high-resolution inputs divides the Transformer blocks into $N$ subsets, each containing $M$ blocks. Within each subset, the first \(M-1\) blocks compute local window attention, while the final block applies global attention to enable cross-window information propagation. Fig.~\ref{fig:mix-res-vitdet} illustrates this organization using ViTDet as an example. We use ViTDet-L~\cite{vitdet-l-git}, initialized from its pre-trained checkpoint, as the concrete model instantiation throughout this paper. Its backbone contains \(N=4\) subsets.

To process mixed-resolution inputs, ViTrail projects low-resolution and full-resolution image patches into token embeddings with the same pre-trained patch-embedding layer. The corresponding global positional embeddings are then added to both sets of tokens before they are concatenated and processed by the Transformer blocks. To recover a regular full-resolution token grid, ViTrail bilinearly upsamples the $w\times w$ token grid of each downsampled region to $wd\times wd$ at a selected stage of the backbone and places it back at the region's position. We refer to this stage as the \emph{restoration point (RP)} and denote it by $\beta$. In principle, the RP can be placed at any depth within the ViT backbone. In practice, we consider one candidate RP per subset, immediately before its global-attention block. The backbone also adds relative positional embeddings in its attention blocks. Window-attention blocks keep them, since each window forms a regular token grid at either resolution. Global-attention blocks before the RP omit them, because tokens of different resolutions do not lie on one regular grid, and rely on the global positional embeddings alone.

These candidate RPs provide a direct way to control how much of the backbone operates on the shorter mixed-resolution sequence. Moving the RP later allows more Transformer blocks to process fewer tokens and therefore increases the potential inference savings. It can also lower inference accuracy, because more blocks then process a token layout that the pre-trained backbone was not trained on. Once restoration is applied, the subsequent blocks operate on a regular full-resolution token grid. We also include $\beta=0$ as a special case in which the downsampled image regions are upsampled to full resolution before patch embedding, so every block processes the full-resolution sequence. In all cases, the restoration mechanism changes only the token layout processed by the backbone and leaves its architecture unchanged.

\section{Runtime Configuration Adaptation}

\subsection{Accuracy-Oriented Region Selection}

Let $\mathcal{R}$ denote the set of all decision regions. For each frame scheduled for offloading, ViTrail determines a binary downsampling mask $\mathbf{B}\in\{0,1\}^{|\mathcal{R}|}$, where $B_j=1$ indicates that decision region $j$ is downsampled. The goal is to reduce E2E offloading latency while preserving the inference fidelity of task-relevant content. Because the appropriate choice of $\mathbf{B}$ depends on the current frame content, the mask must be adapted at runtime. Allowing an independent downsampling decision for every region yields $2^{|\mathcal{R}|}$ possible masks. Exhaustively evaluating this exponentially large mask space is impractical for real-time operation. At decision time, the device estimates the task relevance of the current frame from the latest remote inference result and local tracking. Even an exact solver that met the per-frame deadline would optimize the mask only with respect to these estimates. To bound the per-frame search cost, we consider only masks induced by three downsampling policies defined over region categories.

To construct these policies, we characterize each decision region $j$ using two metrics: \emph{motion} ($m_j$) and \emph{task relevance} ($\rho_j$). Because ViTrail targets moving-camera MVA, captured frames often exhibit substantial apparent motion in both task-relevant and background regions. The Region Motion Analyzer estimates $m_j$ using a lightweight background subtraction model~\cite{cv-bg}. Specifically, $m_j$ is calculated as the ratio of foreground pixels (i.e., pixels exhibiting significant motion) in region $j$ to the total number of foreground pixels in the frame. Thus, $m_j$ quantifies the normalized contribution of region $j$ to the frame-wide detected motion, rather than the absolute magnitude of local motion. The task relevance $\rho_j$ reflects the importance of region $j$ to the downstream task and is derived from the latest remote inference result together with the result currently propagated by the Local Tracker. For object detection, both results consist of bounding boxes, and $\rho_j$ is the fraction of these boxes that overlap region $j$.

We use $m_j$ and $\rho_j$ to partition the decision regions into three operational categories for constructing candidate downsampling masks. These categories support downsampling decisions without requiring the system to explicitly determine the physical source of the observed motion. First, \emph{static background regions (SBRs)} have low task relevance and make a negligible contribution to the frame-wide detected motion, typically corresponding to visually stable background content (e.g., sky or calm water). Second, \emph{camera-induced motion regions (CMRs)} have low task relevance but make a non-negligible contribution to the detected motion. In moving-camera videos, such regions often correspond to background scene content that exhibits apparent motion due to camera motion (e.g., buildings or vegetation). Third, \emph{task-relevant regions (TRRs)} have high task relevance and contain content important to the downstream task, such as pedestrians or vehicles in object detection.

\begin{figure*}[!t]
    \centering
    \includegraphics[width=0.95\linewidth]{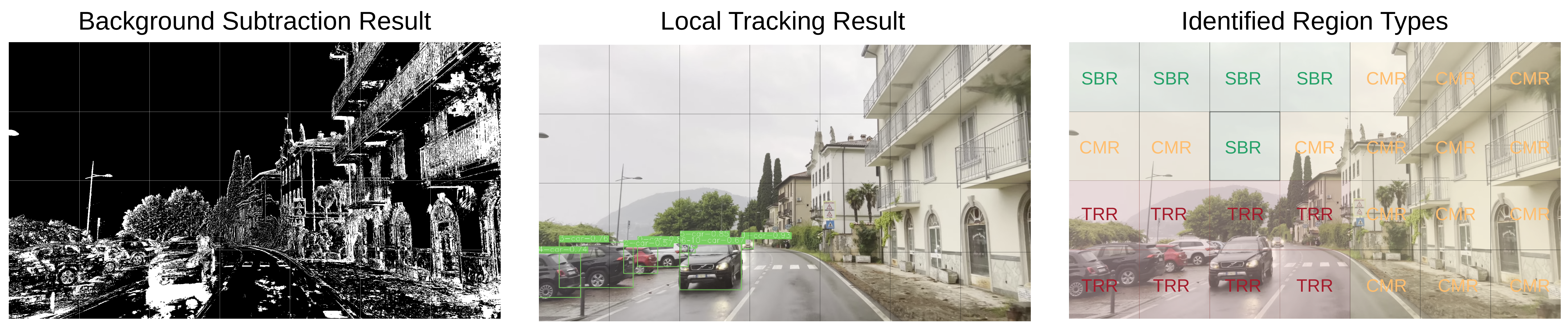}
    \caption{An example of identifying region types based on background subtraction and local tracking.}
    \label{fig:region-types}
    \vspace{-0.3cm}
\end{figure*}

We use a motion threshold $\delta_m$ and a task-relevance threshold $\delta_\rho$ to assign each decision region $j$ to one of these categories. If $\rho_j>\delta_\rho$, the region is classified as a TRR. Among the remaining regions, a region is classified as an SBR if $m_j<\delta_m$ and as a CMR otherwise. This relevance-first rule gives task relevance priority over the subsequent motion-based distinction between SBRs and CMRs. If no object is detected, $\rho_j$ is zero for every region. If no foreground pixel is detected, $m_j$ is undefined, and every region that is not a TRR is classified as a CMR. Fig.~\ref{fig:region-types} illustrates an example of this region-type assignment.

Because downsampling TRRs can cause substantial inference accuracy degradation, ViTrail preserves these regions at full resolution. We define a regional downsampling policy $\tau_d\in\{0,1,2\}$, where each value specifies the region categories targeted for downsampling: no downsampling ($\tau_d=0$), CMRs only ($\tau_d=1$), or both CMRs and SBRs ($\tau_d=2$). The $\tau_d=0$ policy retains the entire frame at full spatial resolution, while the compression quality remains independently controlled by $\lambda$. We do not include a policy that downsamples only SBRs because our offline profiling consistently shows that it provides a less favorable encoded-size--accuracy trade-off than downsampling only CMRs. In our target videos, SBRs are generally compressed efficiently even at full resolution, whereas downsampling CMRs yields larger reductions in encoded frame size at a comparable loss of inference accuracy.

For a selected downsampling policy $\tau_d$, ViTrail sets $B_j=1$ if region $j$ belongs to a category targeted for downsampling by $\tau_d$ and $B_j=0$ otherwise. By construction, $B_j=0$ for all TRRs under every candidate policy. Thus, $\tau_d$ specifies a category-level downsampling policy, while $\mathbf{B}$ is its frame-specific realization. Even for a fixed $\tau_d$, $\mathbf{B}$ varies across frames as the motion and task relevance of the regions change.

\subsection{Offloading Latency Estimation}
\label{sec:latency-est}

Let $N_d = \sum_{j\in\mathcal{R}} B_j$ denote the number of decision regions downsampled by $\mathbf{B}$. On the mobile device, applying $\mathbf{B}$, constructing the mixed-resolution frame, and encoding the resulting frame are all performed before transmission. We jointly account for these operations using the preparation-and-encoding delay $T^{\text{prep+enc}}$. Our offline profiling on the target mobile device shows that this combined delay is well characterized by the number of downsampled regions $N_d$ and the compression quality parameter $\lambda$. We therefore estimate it using the profiled function $\hat{T}^{\text{prep+enc}}(N_d,\lambda)$. Although the frame decoding delay $T^{\text{dec}}$ varies slightly across configurations, its magnitude and variation are small relative to the other latency components. We consequently approximate it using its mean value from offline profiling on the target server, denoted by $\hat{T}^{\text{dec}}$.

After decoding the frame, the server uses $\mathbf{B}$ to construct the mixed-resolution token sequence and executes the vision model. Both operations are included in the inference delay $T^{\text{inf}}$. For a fixed region partition and downsampling factor $d$, $N_d$ determines the initial token sequence length, while $\beta$ determines how many Transformer blocks process the shorter sequence before resolution restoration. Our profiling shows an approximately linear relationship between $N_d$ and $T^{\text{inf}}$ for each $\beta$. We therefore fit a separate linear model for each $\beta$ and estimate the inference delay as $\hat{T}^{\text{inf}}_{\beta}(N_d) =\mathrm{LM}^{\text{inf}}_{\beta}(N_d)$.

We next estimate the uplink payload size. Let $S(\bm{c})$ denote the total amount of data transmitted from the mobile device for configuration $\bm{c}$. The binary mask $\mathbf{B}$ and the restoration-point index $\beta$ introduce only a small metadata overhead, so the variation in $S(\bm{c})$ is dominated by the compressed mixed-resolution frame. At a given $\lambda$, its size is governed by the spatial detail that remains in each region at the resolution at which the region is encoded. We characterize this detail using gradient energy. ViTrail computes the image gradient on the half-resolution frame already produced by the Region Motion Analyzer and sums its magnitude within each decision region. For a candidate mask $\mathbf{B}$, the per-region sums are partitioned into $E^{\text{keep}}(\mathbf{B})$ for regions retained at full resolution and $E^{\text{ds}}(\mathbf{B})$ for downsampled regions. Both terms are computed directly for each candidate mask, and the per-region gradient sums are computed once per frame and shared by all candidates. We train a lightweight multilayer perceptron (MLP) to estimate the uplink payload size as

\begin{equation}
\hat{S}(\bm{c}) =
\mathrm{MLP}^{\text{size}}
\bigl(\lambda,
E^{\text{keep}}(\mathbf{B}),
E^{\text{ds}}(\mathbf{B})\bigr).
\label{equ:size-estimator}
\end{equation}

Let $\hat{U}$ denote the estimated uplink throughput. ViTrail estimates it as the mean over a short window of the most recent offloads, because the throughput of a mobile link changes between consecutive offloads. Accordingly, $\hat{S}(\bm{c})/\hat{U}$ estimates the payload transmission time of the uploaded frame. Because the returned inference result is small relative to the available downlink throughput, its payload transmission time is negligible. Nevertheless, both the frame upload and the result download incur one-way network latency. We approximate their combined network latency using the measured round-trip time (RTT), denoted by $\hat{L}^{\text{net}}$.

Combining the device-side processing, network, decoding, and inference delays, the Performance Estimator predicts the E2E offloading latency under configuration $\bm{c}$ as

\begin{equation}
\hat{T}(\bm{c}) =
\hat{T}^{\text{prep+enc}}(N_d,\lambda)
+ \frac{\hat{S}(\bm{c})}{\hat{U}}
+ \hat{T}^{\text{dec}}
+ \hat{T}^{\text{inf}}_{\beta}(N_d)
+ \hat{L}^{\text{net}}.
\label{equ:offload-latency}
\end{equation}

The offline payload-size and preparation-and-encoding models can exhibit level shifts across scenes and device loads. ViTrail therefore corrects them online. After each offload, it computes multiplicative correction factors from the ratios of the measured payload size and preparation-and-encoding delay of the transmitted configuration to their uncorrected model predictions, and applies them to the corresponding estimates of all candidates in the next decision.

\subsection{Configuration Selection by Installation Accuracy}
\label{sec:config-select}

\begin{figure}[!t]
    \centering
    \includegraphics[width=\linewidth]{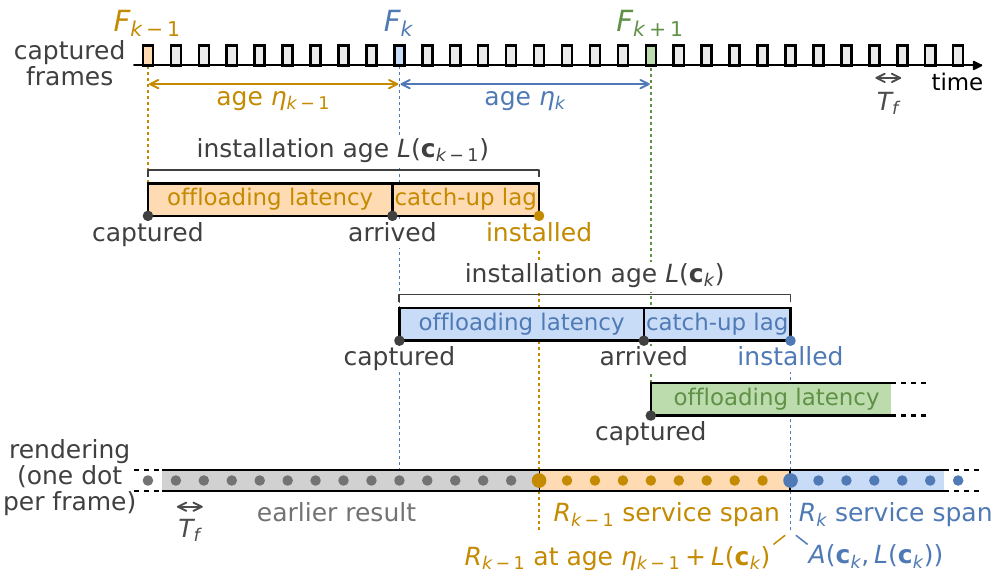}
    \caption{How configuration selection influences rendering accuracy.}
    \label{fig:two-period}
\end{figure}

Configuration selection must account not only for the accuracy of a remote inference result but also for when that result becomes available for rendering. Since each offload typically spans several frame capture intervals of duration $T_f$, a remote result is usually already several frames behind the live stream when it arrives. The Local State Manager therefore first performs \emph{catch-up tracking}, propagating the result forward until it reaches the live frame. We say that a remote result is \emph{installed} when catch-up tracking completes and a tracking result derived from it is first used for rendering. We define the \emph{catch-up lag} $L_{\mathrm{catch}}$ as the number of capture intervals from result arrival to installation (Fig.~\ref{fig:two-period}). We further define the \emph{age} of a remote result at a frame as the number of capture intervals elapsed since its offloaded frame. The age of a remote result when it is installed is its \emph{installation age}, denoted by $L(\bm{c})$ for a result produced under configuration $\bm{c}$.

After installation, a remote result serves subsequent frames until the next remote result is installed. We refer to this interval as its \emph{service span}. Within a service span, each rendered frame uses a tracking result propagated from the serving remote result. Rendering accuracy therefore depends on both the fidelity of the remote inference result and its age, as tracking drift and scene changes accumulate over time. We capture this dependence with $A(\bm{c},\ell)$, the accuracy of a tracking result at age $\ell$ when the corresponding remote result is produced under configuration $\bm{c}$. Age $\ell=0$ corresponds to the offloaded frame itself, so $A(\bm{c},0)$ is the inference accuracy of the remote result.

We now consider how the configuration of one offload affects rendering accuracy across the transition from one remote result to the next. Index the offloads by $k$, and let $F_k$, $\bm{c}_k$, and $R_k$ denote the offloaded frame, the selected configuration, and the remote result of the $k$th offload, respectively. Here $\bm{c}_k$ is the configuration denoted $\bm{c}^\star$ in the system overview. As Fig.~\ref{fig:two-period} shows, $F_k$ is captured after $R_{k-1}$ arrives, while $R_{k-1}$ may still be undergoing catch-up tracking. Let $\eta_{k-1}$ denote the age of $R_{k-1}$ when $F_k$ is captured, that is, the number of capture intervals between $F_{k-1}$ and $F_k$. Result $R_k$ is installed $L(\bm{c}_k)$ capture intervals after $F_k$ is captured, which ends the service span of $R_{k-1}$ and begins that of $R_k$. By then, $R_{k-1}$ has reached age $\eta_{k-1}+L(\bm{c}_k)$, while $R_k$ begins serving at age $L(\bm{c}_k)$ with accuracy $A(\bm{c}_k,L(\bm{c}_k))$, which we call its \emph{installation accuracy}. The choice of $\bm{c}_k$ therefore determines both how old $R_{k-1}$ becomes before its service span ends and how accurate $R_k$ is when its service span begins.

To prevent a serving result from becoming too stale, the MVA application specifies a freshness target $\ell_{\max}$ that bounds the age of a remote result at the end of its service span. When the configuration of the $k$th offload is selected, the installation age and accuracy of each candidate are not yet known, so the Performance Estimator predicts them. Let $\hat{L}(\bm{c})$ denote the predicted installation age of candidate $\bm{c}$ and $\hat{A}(\bm{c},\ell)$ its predicted accuracy at age $\ell$. A candidate $\bm{c}$ is feasible only if $\eta_{k-1}+\hat{L}(\bm{c})\leq\ell_{\max}$. Among the feasible candidates, the Offload Optimizer selects the one with the highest predicted installation accuracy,
\begin{equation}
\bm{c}_{k}
=
\arg\max_{\bm{c}\in\mathcal{C}}
\hat{A}(\bm{c},\hat{L}(\bm{c}))
\quad
\text{s.t.}\quad
\eta_{k-1} + \hat{L}(\bm{c}) \leq \ell_{\max},
\label{equ:config-select}
\end{equation}
and falls back to the fastest candidate when none is feasible. A frame with no task-relevant region skips estimation and is offloaded at full resolution with the lowest JPEG quality.

This formulation captures information that simpler objectives and constraints omit. Bounding only the E2E offloading latency $\hat{T}(\bm{c})$ or the predicted installation age $\hat{L}(\bm{c})$ would ignore the age $\eta_{k-1}$ already accumulated since $F_{k-1}$. The freshness constraint instead spans the interval from $F_{k-1}$ to the installation of $R_k$, covering the two consecutive offloading periods as well as the catch-up required before $R_k$ begins serving. Likewise, maximizing only the zero-age accuracy $\hat{A}(\bm{c},0)$ would ignore the accuracy lost to tracking propagation before the new result first serves rendering.

Using the predicted E2E offloading latency $\hat{T}(\bm{c})$ in \cref{equ:offload-latency} and neglecting the short time between frame capture and the start of frame preparation, the Performance Estimator predicts the installation age as
\begin{equation}
\hat{L}(\bm{c}) =
\frac{\hat{T}(\bm{c})}{T_f} + \hat{L}_{\mathrm{catch}},
\label{equ:install-age}
\end{equation}
where $\hat{L}_{\mathrm{catch}}$ is an online estimate of the catch-up lag, computed as the median of the most recent valid observations. Although the catch-up lag can increase with the number of frames accumulated before result arrival, this dependence is weak because catch-up tracking is substantially faster than the frame capture rate. Its variation is instead dominated by video content and device load. We therefore use the recent online estimate for all candidates within a configuration decision. The predicted installation accuracy $\hat{A}(\bm{c},\hat{L}(\bm{c}))$ is obtained from an offline profile of $A(\bm{c},\ell)$ built for every configuration at a discrete set of ages. To construct the profile, sample frames are offloaded under configuration $\bm{c}$, and each returned result is propagated by the Local Tracker, as at runtime, to the frame at age $\ell$. The resulting tracking result is then scored against the reference detections for that frame, obtained by running the same model on the unmodified full-resolution frame. At runtime, the profile is linearly interpolated at $\hat{L}(\bm{c})$.

\section{Evaluation}

\subsection{Prototype Implementation}
\label{sec:prototype}
We deploy ViTrail on an NVIDIA Jetson Orin Nano mobile device~\cite{jetson} and an edge server equipped with an Intel i9-13900K CPU and an NVIDIA GeForce RTX~5090 GPU. The two devices are connected through a TP-Link AX3000 router, with the unshaped path providing nearly 1\,Gbps throughput. To ensure reproducible and fair comparisons, we control both the video input and network conditions. Unless otherwise stated, a GStreamer pipeline~\cite{gstreamer} on the Jetson replays a pre-recorded 1080p video with hardware decoding in real time at 30\,FPS, a representative mobile-camera setting. Meanwhile, Mahimahi~\cite{netravali2015mahimahi} replays a fixed real-world 4G/5G trace on the client--server path. Thus, for each video--trace pair, all evaluated methods receive the same frame stream and experience the same time-varying uplink throughput and network latency. Only the camera input and network conditions are replayed. The encoder, inference model, and tracker all execute in real time on the two devices, so the processing and queueing delays on both are measured rather than simulated.

Both devices run Ubuntu 22.04, and ViTrail is implemented in Python. The size MLP is implemented with PyTorch. OpenCV~\cite{opencv} handles JPEG encoding and decoding, and ZeroMQ~\cite{zeroMQ} transports offloading data and inference results. On the mobile device, the Region Motion Analyzer runs in parallel with local tracking, with background subtraction on the GPU and per-region gradient computation for the size estimator on the CPU. Catch-up tracking upon receiving a remote result is handled asynchronously, and the observed catch-up lag is recorded to update $\hat{L}_{\mathrm{catch}}$. The Local Tracker uses Lucas--Kanade optical flow~\cite{lucas1981iterative} for real-time tracking and low-latency catch-up tracking on the mobile device.

\begin{table}[t]
    \centering
    \caption{Videos used in the evaluation.}
    \label{tab:videoset}
    \small
    \setlength{\tabcolsep}{5pt}
    \begin{tabular}{llll}
        \toprule
        \textbf{Video ID} &
        \textbf{Scenario and source} &
        \textbf{Capture condition} &
        \textbf{Motion} \\
        \midrule
        walkS  & Countryside walk~\cite{video-walkS} & Sunny     & Low  \\
        walkR  & City walk~\cite{video-walkR}        & Rainy     & Low  \\
        walkB  & Indoor walk, self-shot       & Artificial lighting
        & Low \\
        cycleS & Cycling~\cite{video-cycleS}          & Sunny     & High \\
        driveN & Night drive~\cite{video-driveN}       & Night     & High \\
        \bottomrule
    \end{tabular}
\end{table}

\subsection{Evaluation Setup}
\label{sec:eval-setup}

\textbf{Video dataset.} Table~\ref{tab:videoset} summarizes the videos used for evaluation. The dataset covers five moving-camera scenarios with different motion levels, indoor and outdoor environments, and lighting and weather conditions. The videos are either sourced from YouTube~\cite{video-walkS,video-walkR,video-cycleS,video-driveN} or recorded by us (\emph{walkB}) using a body-worn Insta360 GO~3S camera~\cite{insta360-go3s}. From each video source, we extract two disjoint clips, a 60-second clip for estimator training and profiling and a 300-second clip for evaluation, yielding 45,000 evaluation frames in total. We adopt a \emph{leave-one-video-out} evaluation protocol for ViTrail and all baselines. For each held-out video, all content-dependent profiles and estimators, including the baselines', are built using only the training clips of the other four videos.

\textbf{Network traces.} We collect 60 real-world mobile network traces from multiple public sources~\cite{cellular_ghoshal2022depth, cellular_imc_ghoshal2023performance, raca2018_beyond_throughput}, evenly divided between 4G and 5G. Each trace spans 300 seconds and provides network measurements at a 1-second resolution. The per-trace mean uplink throughput ranges from 6.4 to 36.4\,Mbps for 4G and from 4.8 to 135.5\,Mbps for 5G. The mean RTTs of the 4G and 5G traces are 38 and 32\,ms, respectively. Pairing each video with all 60 network traces yields 300 video--trace evaluation pairs.

\textbf{Server-side inference model.} We use ViTDet-L~\cite{vitdet-l-git} for server-side inference. ViTrail's region partitioning and resolution restoration apply to any window size, including the native $14\times14$ windows of the original checkpoint. In the high-resolution videos we target, most objects are much smaller than the decision regions induced by the native windows. Preserving detail in these objects can therefore require retaining substantial surrounding background at full resolution. Our prototype uses $9\times9$ windows to allow more selective regional downsampling. With $d=2$, this increases the number of decision regions per padded 1080p frame from 15 to 28. We fine-tune the checkpoint offline on COCO~\cite{lin2014microsoft} for the smaller windows, and all methods use the resulting model. The resulting model achieves 54.7 box AP on COCO val2017, compared with 55.5 reported for the original checkpoint~\cite{vitdet-l-git}. Later restoration points reduce inference delay but lower accuracy, as discussed in \cref{sec:restore-pos}. To mitigate this loss, our prototype adds a LoRA adapter~\cite{hu2022lora} that is active only when $\beta\geq3$. The adapter has rank 16 and is applied to the query-key-value and output projections of all attention blocks, with about 2.4\,M parameters. It is trained offline by self-distillation, with the model's own output at $\beta=0$ as the target, so it needs no labels. It adds less than 1\,ms to inference.

\textbf{Configuration and hyperparameter settings.} The configuration space includes three regional downsampling policies $\tau_d \in \{0,1,2\}$, five JPEG quality values $\lambda \in \Lambda = \{75,80,85,90,95\}$, and five restoration points $\beta \in \{0,1,2,3,4\}$. When $\tau_d=0$, we fix $\beta=0$ because no low-resolution tokens require restoration, yielding 55 candidate configurations. Based on offline profiling, we set the motion threshold $\delta_m$ to 0.005. We set the task-relevance threshold $\delta_\rho$ to 0 so that every region with nonzero task relevance is classified as a TRR. ViTrail's freshness target $\ell_{\max}$ is 30 capture intervals, equivalent to 1\,s at 30\,FPS. The uplink throughput $\hat{U}$ and network latency $\hat{L}^{\text{net}}$ are averaged over the two most recent offloads. We initialize $\hat{L}_{\mathrm{catch}}$ to 2 capture intervals and update it using the median of the 64 most recent valid catch-up observations. The installation-accuracy profile covers $\ell \in \{0,6,9,12,16,20,24\}$. Before profile interpolation, predicted installation ages are clamped to the range from 6 to 24 capture intervals, while the freshness constraint uses the unclamped predictions. $\mathrm{MLP}^{\mathrm{size}}$ consists of three fully connected layers with 64, 64, and 1 neurons, respectively.

\textbf{Performance metrics.} Following prior systems~\cite{liu2019edge,du2020server,zhang2022casva}, we obtain reference detections by running the same fine-tuned ViTDet-L model on every unmodified full-resolution frame. These reference detections isolate degradation introduced by the offloading pipeline from the detector's intrinsic error and represent an idealized setting in which full-resolution inference is immediately available for every frame. Both accuracy metrics below are micro-averaged F1 scores of result boxes against these reference detections. A result box and a reference detection match when they have the same class and their IoU exceeds 0.5. Matches are made one to one in decreasing order of the result boxes' confidence. We report the following metrics.

\begin{itemize}[leftmargin=*]
    \item[$\circ$] \emph{Rendering accuracy.} The F1 score of the final rendered results over all frames of a video--trace pair. As in prior edge-assisted systems~\cite{liu2019edge, kong2023accumo, kong2024arise}, it scores the result used for every frame, including frames rendered from tracked results, rather than only the offloaded frames. This is our primary application-oriented quality metric and captures the joint effects of inference fidelity and result freshness.

    \item[$\circ$] \emph{Inference accuracy.} The F1 score of the server-side inference results over all offloaded frames of a video--trace pair. It measures the fidelity degradation introduced by compression and mixed-resolution processing.

    \item[$\circ$] \emph{E2E offloading latency.} The measured counterpart of $\hat{T}(\bm{c})$ in \cref{equ:offload-latency}, from the start of device-side frame preparation to reception of the remote result. It drives result freshness.
\end{itemize}

\begin{figure*}[!t]
\centering
\includegraphics[width=\linewidth]{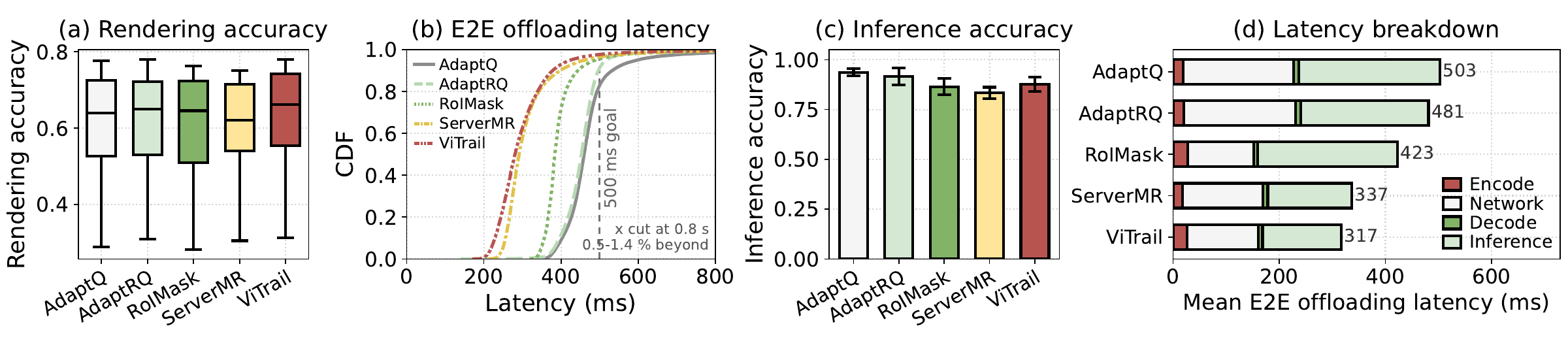}
\caption{Overall performance over the 300 video--trace pairs, with (a)~rendering accuracy per pair, (b)~cumulative distribution function (CDF) of E2E offloading latency over all offloads, (c)~inference accuracy as mean and standard deviation over the pairs, and (d)~mean E2E offloading latency by component.}
\label{fig:overall-v2}
\end{figure*}

\textbf{Baselines.} All baselines augment the canonical edge-assisted MVA paradigm~\cite{chen2015glimpse, kong2024arise} with the same Local Tracker and Renderer as ViTrail, and differ in how they reduce or adapt transmission and inference costs.

\begin{itemize}[leftmargin=*]

    \item[$\circ$] \underline{\emph{AdaptQ.}} A latency-adaptive transmission baseline inspired by adaptive video streaming~\cite{mao2017neural, fang2024robust}. It offloads full-resolution frames and adapts only the JPEG quality $\lambda \in \Lambda$. It predicts the E2E offloading latency of each candidate configuration with \cref{equ:offload-latency}, using an offline size profile with an online size correction and the same network observations as ViTrail. Among the candidate configurations that meet a latency goal, it selects the one with the highest offline-profiled inference accuracy, and it falls back to the fastest one when none does. Since the adaptive baselines do not explicitly model result installation or catch-up, we set their fixed per-offload latency goal to $\ell_{\max} T_f / 2 = 500$\,ms, a nominal equal split of ViTrail's freshness target (\cref{sec:config-select}) across two consecutive offloads.

    \item[$\circ$] \underline{\emph{AdaptRQ.}} A latency-adaptive baseline for joint transmission and inference optimization, inspired by configuration-adaptive video analytics~\cite{zhang2022casva}. It extends AdaptQ by jointly selecting a uniform spatial scale and JPEG quality from $\{1,0.75,0.5\} \times \Lambda$, using offline profiles of payload size and inference accuracy under the same size correction, latency goal, and selection rule. The server performs inference at the selected resolution, so downsampling reduces both the transmitted payload and the ViT token count. It serves as the uniform-resolution baseline.

    \item[$\circ$] \underline{\emph{RoIMask.}} A content-aware transmission baseline representing RoI-based compression strategies~\cite{du2020server, liu2022adamask}. It masks decision regions not classified as TRRs and offloads the resulting frame at full resolution. We evaluate all JPEG quality values in $\Lambda$ in a preliminary sweep and report results using the fixed value ($\lambda=75$) that maximizes mean rendering accuracy.

    \item[$\circ$] \underline{\emph{ServerMR.}} A content-aware inference baseline following server-side token reduction methods~\cite{rao2021dynamicvit, bolya2023token, ronen2023vision}. The device offloads the full-resolution frame, after which the server ranks the same decision regions used by ViTrail with the feature-based saliency scorer from~\cite{ronen2023vision}, downsamples the least salient regions, and performs mixed-resolution inference with restoration at $\beta=4$. We fix the number of downsampled regions to the mean number of decision regions without a task-relevant object in the training clips. JPEG quality is swept and selected as in RoIMask, which also yields $\lambda=75$. Because it restores at $\beta=4$, ServerMR uses the same server-side inference model as ViTrail, including the adapter.

\end{itemize}

\subsection{End-to-End Performance}
\label{sec:e2e-performance}

\begin{table}[!t]
\caption{Mean rendering accuracy per video and over all 300 pairs, best in bold.}
\label{tab:per-video}
\centering
\small
\begin{tabular}{l ccccc}
\toprule
Video & AdaptQ & AdaptRQ & RoIMask & ServerMR & ViTrail \\
\midrule
walkS & 0.730 & 0.725 & 0.720 & 0.701 & \textbf{0.737} \\
walkR & 0.636 & 0.642 & 0.637 & 0.621 & \textbf{0.661} \\
walkB & 0.528 & 0.534 & 0.519 & \textbf{0.558} & 0.555 \\
cycleS & 0.692 & 0.699 & 0.697 & 0.696 & \textbf{0.716} \\
driveN & 0.502 & 0.504 & 0.487 & 0.514 & \textbf{0.529} \\
\midrule
All & 0.618 & 0.621 & 0.612 & 0.618 & \textbf{0.640} \\
\bottomrule
\end{tabular}
\vspace{-0.5cm}
\end{table}

\textbf{Rendering accuracy.} ViTrail achieves the highest mean rendering accuracy and the lowest median E2E offloading latency among all methods. Table~\ref{tab:per-video} shows a mean rendering accuracy of 0.640 across the 300 video--trace pairs, compared with 0.612 to 0.621 for the baselines. ViTrail outperforms each baseline on 81\% (ServerMR) to 95\% (AdaptQ) of the pairs and leads on four of five videos, trailing only ServerMR on walkB.

Fig.~\ref{fig:overall-v2}(a)--(c) show that high inference accuracy alone does not guarantee high rendering accuracy, because a longer offloading latency makes a result older when it serves rendering. AdaptQ and AdaptRQ select configurations by inference accuracy under a latency goal and retain the highest inference accuracies, 0.937 and 0.915, but at median offloading latencies of 460 and 453\,ms, respectively. RoIMask and ServerMR have lower median latencies of 384 and 291\,ms, respectively, but also lower inference accuracies of 0.865 and 0.833. Despite these different accuracy--latency trade-offs, the methods achieve similar rendering accuracy. With a shared mixed-resolution representation and installation-accuracy-driven configuration selection, ViTrail reduces median offloading latency to 281\,ms, 38.0\% below AdaptRQ, the baseline with the highest mean rendering accuracy, while retaining an inference accuracy of 0.877, higher than both RoIMask and ServerMR. It also completes 911 offloads per pair on average, compared with 580 to 851 for the baselines, so its remote results are refreshed more often.

\textbf{Latency breakdown.} Fig.~\ref{fig:overall-v2}(d) decomposes the mean E2E offloading latency over all offloads into device-side preparation and encoding, server-side decoding and inference, and network delay computed as the remainder. Against AdaptQ, AdaptRQ, and RoIMask, the largest latency savings come from inference. AdaptQ and RoIMask process all offloaded frames at full spatial resolution and require 263 to 265\,ms for inference. AdaptRQ averages 241\,ms with occasional whole-frame downsampling. ViTrail instead downsamples an average of 19 of the 28 decision regions per offload and reduces inference delay to 148\,ms, about 44\% below full-resolution inference. Although RoIMask achieves the smallest payload of 100\,KiB per frame (Table~\ref{tab:uplink}) and the lowest network delay of 124\,ms, its full-resolution inference limits the E2E benefit of transmission optimization alone. The two adaptive baselines incur the largest network delays, 208 to 210\,ms, against 134\,ms for ViTrail.

The comparison with ServerMR highlights the benefit of preparing the shared mixed-resolution representation before transmission. ServerMR downsamples about as many regions as ViTrail and executes the ViT in 147\,ms, but server-side saliency scoring and mixed-resolution frame construction add 11\,ms, bringing its inference delay to 158\,ms. It also transmits full-resolution frames with a mean payload of 196\,KiB, compared with 170\,KiB for ViTrail. The corresponding network delays are 152 and 134\,ms, respectively. ViTrail thus retains the inference savings of mixed-resolution processing while also reducing the transmitted payload. Constructing and encoding ViTrail's mixed-resolution frame takes 26\,ms on average, compared with 18 to 19\,ms for AdaptQ and ServerMR. The combined inference and network savings more than offset this cost, giving ViTrail the lowest mean E2E offloading latency.

\begin{figure*}[!t]
    \begin{minipage}[t]{0.33\linewidth}
        \centering
        \includegraphics[width=\linewidth]{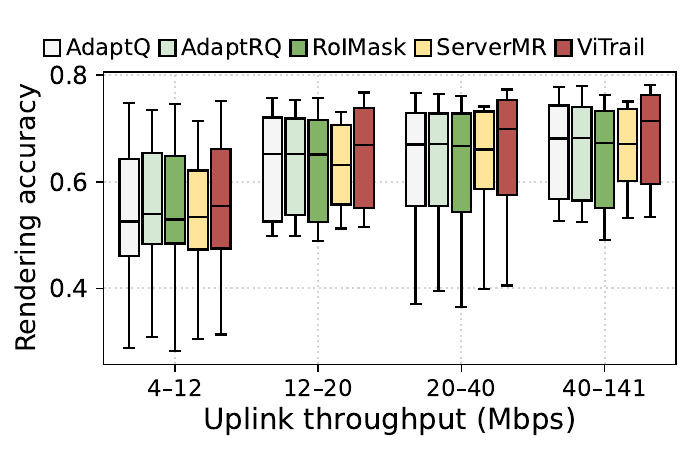}
        \caption{Uplink throughput sensitivity.}
        \label{fig:strata}
    \end{minipage}\hfill\begin{minipage}[t]{0.33\linewidth}
        \centering
        \includegraphics[width=\linewidth]{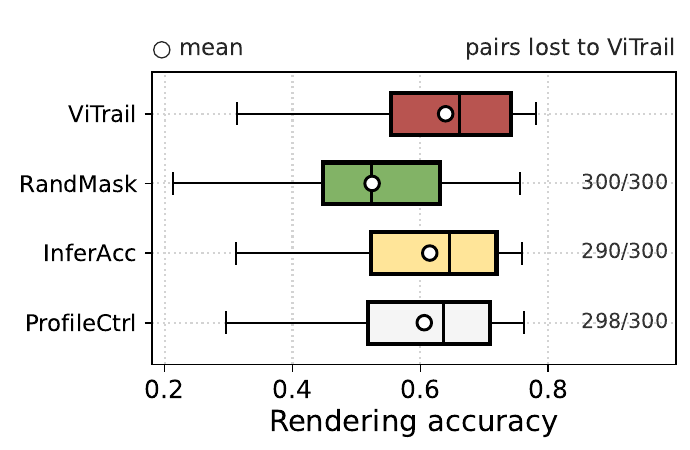}
        \caption{Ablation study.}
        \label{fig:ablation}
    \end{minipage}\hfill\begin{minipage}[t]{0.33\linewidth}
        \centering
        \includegraphics[width=\linewidth]{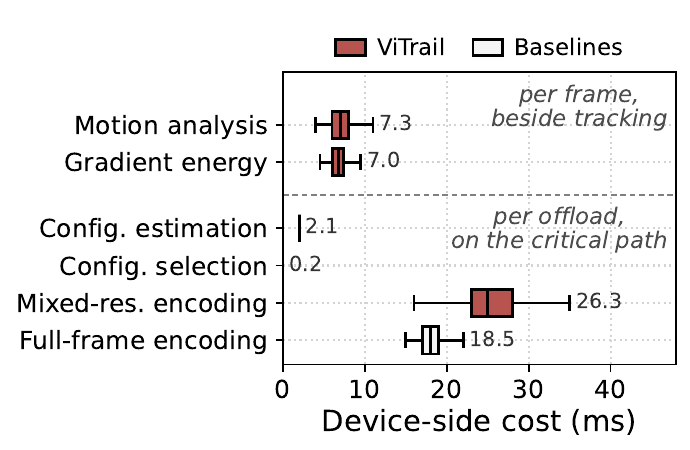}
        \caption{Runtime overhead.}
        \label{fig:overhead}
    \end{minipage}
\end{figure*}

\textbf{Uplink traffic and result rate.} Table~\ref{tab:uplink} reports each method's mean uplink data rate and result rate. For each video--trace pair, we divide the bits of the encoded frames it sends and the number of inference results it receives by the 300\,s duration of the evaluation clip. ViTrail sends 4.24\,Mbps, less than the 5.50 and 5.66\,Mbps of the two adaptive baselines, while returning 3.04 results per second against 1.93 and 2.02. RoIMask sends the least because it masks regions not classified as TRRs, but each frame still passes through full-resolution inference, so it returns only 2.26 results per second.

\begin{table}[t]
\caption{Mean uplink data rate and result rate over the 300 pairs, and mean payload per offloaded frame.}
\label{tab:uplink}
\centering
\small
\setlength{\tabcolsep}{3pt}
\begin{tabular}{l ccc}
\toprule
Method & Uplink (Mbps) & Results/s & KiB/frame \\
\midrule
AdaptQ   & 5.50 & 1.93 & 347 \\
AdaptRQ  & 5.66 & 2.02 & 343 \\
RoIMask  & 1.86 & 2.26 & 100 \\
ServerMR & 4.55 & 2.84 & 196 \\
ViTrail  & 4.24 & 3.04 & 170 \\
\bottomrule
\end{tabular}
\end{table}

\textbf{Network sensitivity.} Fig.~\ref{fig:strata} groups the 300 video--trace pairs into four ranges according to the mean uplink throughput of their network traces, containing 65, 75, 125, and 35 pairs. Median rendering accuracy increases with throughput for every method, and ViTrail has the highest median in every range, rising from 0.554 below 12\,Mbps to 0.713 above 40\,Mbps. ViTrail outperforms each baseline on 74\% to 100\% of the pairs in each range.

The baselines respond to uplink throughput differently. Below 12\,Mbps, AdaptRQ comes closest in rendering accuracy, as it downsamples 57\% of its offloaded frames, cutting its mean E2E offloading latency to 587\,ms against 686\,ms for AdaptQ. Above 20\,Mbps, AdaptRQ returns almost entirely to full resolution, and both adaptive baselines select the highest JPEG quality, $\lambda=95$, on 60\% of their offloads between 20 and 40\,Mbps and on 96\% above 40\,Mbps. Their mean E2E offloading latency remains between 409 and 464\,ms, with full-resolution inference alone taking about 265\,ms. RoIMask's masking saves less transmission time at higher throughput, while its full-resolution inference delay does not shrink. Above 40\,Mbps, ViTrail outperforms it on every pair. ServerMR has an inference delay close to ViTrail's and is the baseline that wins the most pairs against ViTrail above 12\,Mbps.

\subsection{Effectiveness of Key Design Choices}
\label{sec:ablation}

We first evaluate region selection and runtime configuration selection using three variants on the same 300 video--trace pairs, and then the accuracy of the estimators that drive them. Fig.~\ref{fig:ablation} shows each variant's rendering accuracy and the number of pairs it loses to ViTrail.

\textbf{Accuracy-oriented region selection.} \emph{RandMask} replaces the region-type policy $\tau_d$ with the number of regions to downsample as a configuration dimension. It samples that many decision regions uniformly at random and selects configurations using an accuracy profile built with random masks. It loses to full ViTrail on all 300 pairs, and its mean rendering accuracy is 18\% lower than ViTrail's. The lower profiled accuracy of random downsampling leads RandMask to retain full resolution on 55\% of its offloads. On the remaining offloads, its randomly chosen regions can include TRRs, which ViTrail always keeps at full resolution. It averages 684 offloads per pair at a mean E2E latency of 416\,ms, compared with 911 offloads and 317\,ms for ViTrail.

\textbf{Installation-accuracy-driven selection.} \emph{InferAcc} keeps ViTrail's latency estimator, freshness constraint, and fallback, but maximizes the inference accuracy $\hat{A}(\bm{c},0)$ instead of the predicted installation accuracy $\hat{A}(\bm{c},\hat{L}(\bm{c}))$ in \cref{equ:config-select}. Because higher-resolution and higher-quality candidates generally have higher profiled inference accuracy, InferAcc tends to favor slower candidates near the freshness boundary. It offloads at full resolution on 21\% of its offloads and falls back to the fastest candidate more often than ViTrail. InferAcc averages 666 offloads per pair at a mean E2E latency of 427\,ms. It loses to full ViTrail on 290 of the 300 pairs and on all five videos, and its mean rendering accuracy is 4\% lower than ViTrail's.

\emph{ProfileCtrl} further applies the adaptive baselines' profile-driven selection rule to ViTrail's configuration space. It estimates payload size from the offline mean of the training clips with the online corrections in \cref{sec:latency-est}, and selects the candidate with the highest profiled inference accuracy among those predicted to meet the 500\,ms latency goal, instead of applying ViTrail's freshness constraint. ProfileCtrl averages 607 offloads per pair at a mean E2E latency of 483\,ms. It loses to full ViTrail on 298 of the 300 pairs, and its mean rendering accuracy is 5\% lower than ViTrail's.

\textbf{Estimator accuracy.} Under the same leave-one-video-out protocol and before online correction, $\mathrm{MLP}^{\mathrm{size}}$ predicts the encoded frame size on the held-out evaluation clips with a mean absolute percentage error of 6.6\%, compared with 29.7\% for an offline profile that uses the mean size of each configuration in the training clips. With the online correction described in \cref{sec:latency-est}, the error of $\mathrm{MLP}^{\mathrm{size}}$ further decreases to about 4\%. At runtime, the predicted E2E offloading latency of each transmitted configuration has a mean absolute percentage error of 5.5\% relative to its measured latency across the 300 video--trace pairs. Overall, 91\% of the predictions are within 10\% of the measured latency.

\subsection{Runtime Overhead and Behavior}
\label{sec:overhead}

Fig.~\ref{fig:overhead} separates the per-frame costs of motion analysis and gradient computation from the additional costs incurred on offloaded frames. The Region Motion Analyzer runs concurrently with the Local Tracker, with background subtraction and per-region gradient computation taking 7.3\,ms on the GPU and 7.0\,ms on the CPU per frame, respectively. The analyzer and the tracker are threads of the same Python process. Both do most of their work in OpenCV calls, which release the global interpreter lock, so the two can run in parallel. With the analyzer running, tracking takes 14.5\,ms per update on average, and 97.5\% of updates take less than 33\,ms. Estimating the latency and installation accuracy of all 55 candidates takes 2.1\,ms per decision, and configuration selection adds 0.2\,ms. Configuration estimation and selection finish before frame preparation starts and are not part of the E2E offloading latency, while frame construction and encoding are included in the latency breakdown in \cref{sec:e2e-performance}.

\subsection{Sensitivity Analysis}
\label{sec:sensitivity}

\begin{table}[t]
\caption{Results at 30 and 15\,FPS over the same 300 pairs and reference frames.}
\label{tab:fps}
\centering
\small
\setlength{\tabcolsep}{4pt}
\begin{tabular}{l cc cc}
\toprule
 & \multicolumn{2}{c}{Accuracy} & \multicolumn{2}{c}{Offloads per pair} \\
Method & 30 & 15 & 30 & 15 \\
\midrule
ViTrail  & 0.639 & 0.646 & 911 & 873 \\
AdaptRQ  & 0.620 & 0.620 & 605 & 586 \\
ServerMR & 0.618 & 0.617 & 851 & 811 \\
\bottomrule
\end{tabular}
\end{table}

\textbf{Frame rate.} Neither ViTrail nor the baselines offload at a fixed frame interval. A new offload starts shortly after the previous result returns, so offloads run back to back and their frequency depends mainly on how long each offload takes. The capture rate can affect offloading in two ways. First, the delay before a new offload starts can be up to one capture interval, and this bound shrinks as the capture rate rises. Second, the capture rate sets the interval between local tracking updates, whose results guide region selection in ViTrail. In an offline test, updating the same tracker at 15, 30, or 60\,FPS changes its tracking accuracy at the same elapsed time since the offloaded frame by about 0.01 at most. We therefore expect the capture rate to have little effect on region selection.

Table~\ref{tab:fps} compares ViTrail, AdaptRQ, and ServerMR at 15 and 30\,FPS on the same 300 video--trace pairs. The 15\,FPS input keeps every other frame of each 30\,FPS clip. To make the two rates comparable, the 30\,FPS accuracies are scored only on these frames and can therefore differ slightly from those in Table~\ref{tab:per-video}. At 15\,FPS, ViTrail keeps its freshness target of 1\,s and reads its accuracy profiles at the same elapsed times as at 30\,FPS, and AdaptRQ keeps its 500\,ms latency goal. We compare against these two baselines because AdaptRQ has the highest mean rendering accuracy among the baselines and ServerMR outperforms ViTrail on the most pairs. The effect on offloading is small. Halving the capture rate reduces the number of completed offloads by only 3\% to 5\% for each method, and the median E2E offloading latency changes by less than 3\,ms. The drop in offloads is consistent with the longer wait before each offload at 15\,FPS. ViTrail achieves the highest mean rendering accuracy among the three methods at both rates. At 15\,FPS, it outperforms AdaptRQ on 286 of the 300 pairs and ServerMR on 294.

\textbf{Resolution.} ViTrail targets high-resolution input, so resolutions below 1080p, such as 720p, are outside the scope of this work. We also run E2E experiments at 1440p with the same three methods at 15\,FPS, using the same server-side model and every profile rebuilt from the 1440p training clips. At this resolution, ViTrail also selects a whole-frame scale before regional downsampling, choosing between native and 0.75, two of the three scales available to AdaptRQ. With this configuration, ViTrail reaches a rendering accuracy of 0.593 over all 300 video--trace pairs, against 0.580 for AdaptRQ and 0.544 for ServerMR, and it leads AdaptRQ in 283 pairs and ServerMR in 299. It completes 770 offloads per pair at a mean E2E offloading latency of 361\,ms, against 579 offloads at 486\,ms for AdaptRQ and 505 at 563\,ms for ServerMR. Its inference accuracy of 0.784 is level with the 0.785 of AdaptRQ and below the 0.807 of ServerMR, so, as at 1080p, its lead is consistent with fresher results. Each resolution is scored against its own reference detections, so accuracies do not compare across resolutions.

\textbf{Network throughput estimation.} Many methods predict mobile network throughput, in the downlink~\cite{lv2022lumos, narayanan2020lumos5g} and the uplink~\cite{lee2020perceive}, and such a predictor could supply the uplink throughput estimate that ViTrail's latency model takes as input. Because throughput prediction is not a contribution of this work, our prototype uses a simple estimator, the mean uplink throughput of the two most recent offloads. We keep the window short because mobile network throughput changes quickly. Our traces provide measurements at a 1-second resolution, while ViTrail completes about three offloads per second, so windows of 8 or 16 offloads span several trace intervals. In a replay of the logged uploads, the two-offload window has lower mean absolute prediction error than windows of 8 or 16 offloads on 58 of the 60 traces.

\textbf{Threshold sensitivity.} For each threshold, we examine which of ViTrail's decisions it can affect. The task-relevance threshold $\delta_\rho$ is set to zero so that every region overlapping a bounding box from detection or tracking is classified as a TRR and retained at full resolution. The motion threshold $\delta_m$ determines which regions $\tau_d=1$ downsamples and leaves the masks of the other two policies unchanged. The freshness target $\ell_{\max}$ is set by the application and affects a decision only when it excludes the candidate the objective would otherwise choose. In a replay of all decisions of ViTrail with their inputs held fixed, raising $\ell_{\max}$ from 1\,s to 2\,s could change at most 7.03\% of them, and lowering it to 0.8\,s could change at most 21.92\%, since a stricter target excludes more candidates.

\textbf{Oracle region selection.} To keep per-frame mask selection tractable, ViTrail uses a lightweight heuristic to generate candidate masks from region types. It identifies TRRs using the latest remote inference result and the Local Tracker. We compare this heuristic with an oracle variant that identifies TRRs from the reference detections of the frame being offloaded, which are unavailable at decision time. Both variants use the same downsampling policies and configuration selection. The oracle thus keeps every region overlapping a reference detection at full resolution. ViTrail's heuristic reaches a mean rendering accuracy of 0.640, against 0.642 for the oracle. By contrast, RandMask, which selects regions at random, has a mean rendering accuracy 18\% lower than ViTrail's (\cref{sec:ablation}). Region selection thus matters, and the heuristic is already close to the oracle.

\section{Discussion}
\label{sec:discussion}

\textbf{Other backbones.} Plain ViT backbones are common in dense prediction. ViTDet~\cite{li2022exploring} and the Segment Anything Model's image encoder~\cite{kirillov2023segment} use single-scale backbones with unshifted attention windows and a few global attention blocks. For this layout, ViTrail sets each region's side to $wd$ patches so that both resolutions yield complete windows. Larger patches or windows therefore give coarser decision regions. With global attention throughout, each side need only span a multiple of $d$ patches, allowing finer partitions. The restoration point depends on which features the neck or head reads. For ViTPose~\cite{xu2022vitpose} and Segmenter~\cite{strudel2021segmenter}, ViTrail could restore the grid at the encoder output. For the plain ViT variants of DPT~\cite{ranftl2021vision}, it could restore the grid once, before the earliest layer whose features the decoder reads. Runtime configuration selection would use latency and accuracy profiles of the new model. For hierarchical backbones, one option is to restore the grid before the first operation that needs regular spatial neighborhoods, which limits token savings to the preceding blocks. We leave extending mixed-resolution processing through the shifted windows and patch merging of Swin~\cite{liu2021swin} or the stage-transition pooling of Hiera~\cite{ryali2023hiera} to future work.

\textbf{Other dense prediction tasks.} ViTrail reduces the cost of the ViT backbone rather than of a task-specific head, so other dense prediction tasks could also benefit from it. For a new task, the task relevance $\rho_j$ would be computed from the task's output instead of bounding boxes, for example from instance masks, from the area that each person's keypoints span, or from the pixels of the classes of interest. The Local Tracker would be adapted to propagate this output, and $\rho_j$ would be estimated from it as for detection. The accuracy profiles would be rebuilt offline with the task's metric, such as mean IoU for semantic segmentation instead of F1. Instance segmentation and pose estimation localize instances as detection does, so they could keep the same region classification. Semantic segmentation labels every pixel, so the saving would depend on how many regions the classes of interest touch.

\section{Conclusion}
We presented ViTrail, a ViT-native MVA system that co-designs frame transmission and ViT inference through a shared mixed-resolution representation, which reduces both transmission and inference costs. ViTrail aligns decision regions with ViT patches and attention windows and restores a regular feature grid at a configurable backbone stage to support dense prediction. At runtime, it jointly selects the regional downsampling policy, compression quality, and restoration point to maximize predicted installation accuracy under a freshness target. In object detection experiments across 300 video--trace pairs, ViTrail achieves higher mean rendering accuracy and lower mean E2E offloading latency than all four baselines.

\IEEEtriggeratref{54}
\bibliographystyle{IEEEtran}
\bibliography{references}

\end{document}